\documentclass[12pt]{article}
\usepackage{latexsym}
\usepackage{psfig}
\usepackage{epsfig}
\usepackage{amsmath}
\usepackage{citesort}
\usepackage{amssymb}
\begin{document}

\begin{center}
{\large \bf The Asymmetric Exclusion Process revisited:
Fluctuations and Dynamics in the Domain Wall Picture}

\vskip 1cm
{ Ludger Santen and C{\'{e}}cile Appert}

\vskip 0.1cm
\small{Laboratoire de Physique Statistique
\footnote{Laboratoire associ\'e aux universit\'es Paris 6,
Paris 7 et au CNRS}\\
Ecole Normale Sup\'erieure, \\
24 rue Lhomond, F-75231 PARIS Cedex 05, France\\
appert@lps.ens.fr, santen@lps.ens.fr}

\date{\today}

\end{center}

\begin{abstract}
We investigate the total asymmetric exclusion process by analyzing the
dynamics of the shock. Within this approach we are able to calculate
the fluctuations
of the number of particles and density profiles not only in the stationary
state but also in the transient regime. We find that the analytical
predictions and the simulation results are in excellent agreement.
\end{abstract}

\vskip 1cm
{\em PACS numbers: 05.40.-a, 05.60.-k, 02.50.Ga}

\vskip 1cm
{\em Key words: asymmetric exclusion process, boundary induced phase
transitions, domain wall theory, transient regime}

\vskip 1cm

\section{Introduction}

The total asymmetric exclusion process (TASEP) is maybe the most
simple lattice model of collective particle transport
\cite{spohn,sz,gunter}. It can be interpreted, e.g., as a basic
model of traffic flow \cite{review} or as a model for biological
transport \cite{MacD69}. In generic steady states, these processes 
are characterized by a non-vanishing mass transport which leads to an
intrinsic non-equilibrium behavior. 
This  is interesting from a
theoretical point of view.  For example, the stationary probabilities of
the system are not given by an equilibrium ensemble. Of particular
interest are systems with open boundary conditions because one can
observe transitions between different bulk states induced by tuning
the boundary conditions \cite{Krug91}.

Another reason for the importance of the TASEP model is the 
possibility to find an exact solution for the stationary state. This has
been done by solving recursion relations for the stationary partition
function \cite{Liggett,Derr92,Gunter93,DEHP}. Later a phenomenological
approach, referred to as domain wall (DW) theory, was proposed \cite{Kolo98}
which was able to reproduce the
exact results for the phase diagram, in the limit of large system
sizes \cite{Belitzky}. The interest of the latter
theory is twofold: First, it leads to an intuitive understanding of
the most important features of the system and second it can be
generalized to more complicated models which are out of scope for an
exact treatment \cite{Popkov99,Popkov00}. 

In our article we test the accuracy of the DW theory for the TASEP
on finite open lattices. In particular we compare the stationary
fluctuations of the number of particles, which are related to the
integrated two-point correlation functions. Beyond that we also
analyze the non-stationary behavior of TASEP, which 
cannot be treated with exact methods.
Rather than investigating the relaxation
spectrum of the process \cite{Dud}, for which numerical evaluation
is limited to very small system sizes ($L \leq 10$),
we focus on the dynamics of directly observable
quantities. We calculated e.g. the time dependent density profiles,
which can be compared easily to simulations (which can be easily
carried out on much larger system sizes). 

The outline of this paper is as follows. In  section 2 we give a
definition of the model and remind briefly of the most important
features of the DW theory.
Then we compare the predictions
for the fluctuations of the number of particles with
simulation results. Finally (section 4), we show how the system responds to a change
of boundary conditions.

\section{The TASEP with open boundaries}
\label{sec_model}

\subsection{Definition of the model} 

The TASEP is defined on a one-dimensional lattice of length $L$. Each
lattice site $(i)$  can be occupied by  
one particle ($\tau_i = 1$) or be empty ($\tau_i = 0$). In
continuous time the dynamics of the 
particles is defined as follows: A pair of sites $(i,i+1)$ is chosen with
probability $dt$, where $dt$ denotes an infinitesimal time-step. If
the site $(i)$ is occupied and the site $(i+1)$ 
is empty one exchanges the positions of particle and hole. All other
local configurations are unchanged. In case of open boundary
conditions one additionally has to define the in- and output of 
particles. At a given time a particle can be introduced on the first
site with probability $\alpha dt$ if the first site is empty. If the
last site of the chain is occupied the particle may escape from  
the chain with probability $\beta dt$. 

The TASEP with continuous time dynamics can be implemented by a random
sequential update, i.e. one first chooses randomly a link between two
sites and then performs the local update. A possible implementation of
the open boundaries is to add two additional sites $0,L+1$ to the
chain, which are occupied with probability $\alpha$ and $1-\beta$
respectively, and act as particle reservoirs.
By definition, one time step includes $L+1$ link choices, and the
corresponding local updates.

For the TASEP it has been shown that the partition function,
i.e. the sum of all stationary weights for a system of size $L+1$, can be
obtained recursively from the results for a system of size $L$
~\cite{Liggett,Derr92}. The recursion relations have been solved
exactly \cite{Gunter93}, e.g. by means of a matrix representation
(MPA) \cite{DEHP}.  
Now we shall discuss how most of these results can be recovered using
the DW theory.

\subsection{Domain wall theory}

The domain wall (DW) theory \cite{Kolo98} gives a phenomenological
description of the system dynamics. The basic idea of this picture is 
that, as long as the entrance (exit) capacity does not exceed the
capacity of the chain, each particle reservoir enforces a domain (of constant density)
in the bulk. At a given time both domains coexist in the chain. The
coexistence of two domains in the chain implies the existence of a
shock, i.e. a region where the two domains meet. In the domain wall
theory one assumes that the shock is sharp, i.e. that it has a finite
width $W$. For a large class of driven lattice gases this requirement
is fulfilled and one has $W \ll L$ already for moderate system sizes $L$. 
After the introduction of the DW theory 
for the TASEP it has also been successfully applied to models
discrete in time and without particle hole symmetry
\cite{Popkov00,Pigorsch}. Therefore the DW 
theory provides a quite general theoretical framework for
models of particle transport.

Below we now cite the ingredients of the DW theory which are of
relevance for the remaining part of the article.
In this theoretical frame,
the dynamics of the shock determines the behavior of
the system. A first characterization of the shock dynamics is possible by means
of the lattice continuity equation 
\begin{equation} 
\frac{\textrm{d}}{\textrm{dt}} \rho(i,t) = j_{i-1}(t) - j_{i}(t), 
\end{equation}
where $j_{i}(t)$ denotes the local particle current at position $i$ and
time $t$ and $\rho(i,t)$ the density at the same site. 
By evaluating this in the continuum limit for the shock position one
finds that the shock moves far from the boundaries with velocity

\begin{equation} 
V = \frac{j_+ -j_-}{\rho_+ - \rho_-},
\label{V-wall}
\end{equation}
where $\rho_+ (\rho_-)$ and $j_- (j_+)$ are respectively the densities
and fluxes 
in the right (left) domains separated by the shock.
 
The motion of the domain wall can be interpreted as a random walk
with hopping rates 
\begin{equation}
D_+ = \frac{j_+}{(\rho_+ - \rho_-)}, \;\;\;\;\; D_- = \frac{j_-}{(\rho_+ - \rho_-)}
\label{defD}
\end{equation}
for a move to the right (left) \cite{Kolo98}. In case of a blocked
entrance or exit 
one can easily verify that the dynamics of the domain wall is
described correctly. In order to generalize this observation, one 
uses that both particle reservoirs are independent.
The interpretation of the shock dynamics as a random walk
allows for an easy calculation of several quantities of interest.

Now we specify the above quantities for the TASEP. An input rate
$\alpha$ leads to a bulk density $\rho_- = \alpha$  and 
an output rate $\beta$ to $\rho_+ = 1 - \beta$.
Inside each domain one
obtains the same behavior as for the periodic system, e.g. one gets
$j = \rho (1-\rho)$ for the flow. 

In case of the TASEP
one obtains still good results if one assumes that one can identify a
single link as the position of the shock. The link is labeled $i$ if
localized between sites $i$ and $i+1$. Thus the wall location varies
from $0$ to $L$ for a lattice of $L$ sites.
On a finite system of size $L$ the  domain wall
performs a random walk in a lattice with reflecting boundary
conditions. Therefore the probability to find the domain wall at
time $t$ and position $i$ can be evaluated from: 
  
\begin{eqnarray}
\frac{d P(i,t)}{dt} & = & D_+ P(i-1,t) + D_- P(i+1,t)\nonumber\\
 & &  - (D_+ + D_-) P(i,t),
\label{eq_D1}
\end{eqnarray}
for $1\leq i \leq  (L-1)$. At the (reflecting) boundaries one has
\begin{eqnarray}
\frac{d P(0,t)}{dt} & = & D_- P(1,t) - D_+ P(0,t) \\
\frac{d P(L,t)}{dt} & = & D_+ P(L-1,t) - D_- P(L,t),
\label{eq_D3}
\end{eqnarray}
where
$D_+ = \frac{\beta(1-\beta)}{1-\alpha-\beta}$
and
$D_- = \frac{\alpha(1-\alpha)}{1-\alpha-\beta}$.

The stationary solution of eq.~(\ref{eq_D1}-\ref{eq_D3}) in the low
density phase 
$\alpha < \beta, \beta \leq 0.5$ is given by $P(i) =
\exp(-(L-x)/\xi)/\mathcal{N}$, where the localization length is given by 
$\xi = \log(D_+/D_-)$ and the normalization $\mathcal{N}$ by
${\mathcal N} = (1-\exp(-(L+1)/\xi))/(1-\exp(-1/\xi))$. In the high
density phase one obtains analogous  
results by using the particle hole symmetry of the TASEP. By using
choice (\ref{defD}) for the hopping rates, one recovers the exact result of the
localization length $\xi$.

\section{Particle number fluctuations in the domain wall picture}

The fluctuations of the number of particles $N =
\sum_{i} \tau_i$ are defined as: 
\begin{equation}
\begin{split}
 \frac{\langle \Delta N \rangle}{L} 
        &=\frac{\langle N^2 \rangle - \langle N \rangle^2}{L} \\
        &=\frac{1}{L} \left [ 2 \sum_{i<j} \langle \tau_i \tau_j \rangle +
          \langle N \rangle - \langle N \rangle^2 \right ] ,
\end{split}
\label{fluct_def}     
\end{equation}
where the brackets  $\langle \dots \rangle$ denote an average over the
stationary ensemble. 
Eq.~(\ref{fluct_def}) illustrates the connection between the
fluctuations and the two-point correlation functions. The fluctuations
of the particle number, as well as the complete large deviation
function, can be calculated using the exact stationary
solution of the TASEP \cite{Derrida}. We show now, as a sensitive test
of the accuracy of the DW theory for finite systems, how the results
of the domain wall theory compare to the simulation results.

\subsection{Particle number fluctuations and the domain wall theory}

We now proceed to calculate the relevant quantities in the framework
of the DW theory. The first quantity one has to calculate is the
average number of particles. In the low density phase the stationary
probability to find the shock at link $i$ is given by $P(i) =
(1/{\mathcal{N}}) \exp(-(L-i)/\xi)$. In the domain wall picture this
means that one finds $(L-i) \rho_+ + i \rho_-$ particles
in the systems. So by summing over all possible shock positions one
gets 
 
\begin{equation}
\begin{split}
 \langle  N \rangle  & = \frac{1}{{\mathcal{N}}} \sum_{i=0}^{L} 
        e^{-(L-i)/\xi} \left [ (L-i) \rho_+ + i  \rho_- \right ] \\
    & =  \rho_- L \\&+ \frac{\delta\,e^{-1/\xi}\,\left( 1 - \left( L + 1\right) \,e^{-L/\xi} + L\,e^{-(L+1)/\xi} \right) }
   {\left( 1 - e^{-1/\xi} \right) \,\left( 1 - e^{-(L+1)/\xi} \right) },
\end{split}
\label{Nmoy}
\end{equation}
where $\delta =\rho_+ -  \rho_-$ denotes the density difference
between the high and low density domain which coexist.
 
Next we have to calculate the two point correlations $\langle \tau_i
\tau_j \rangle$ for $i<j$. This is done by averaging
over the particle densities if the domain wall is located left from 
site $i$, between $i$ and $j$, and right from site $j$, i.e. one finds 
\begin{equation}
\begin{split}
\langle \tau_i\tau_j \rangle & = \frac{\rho_-^2}{\mathcal{N}} \sum_{k=0}^{i-1}
e^{-(L-k)/\xi} +
\frac{\rho_-\rho_+}{\mathcal{N}}\sum_{k=i}^{j-1}e^{-(L-k)/\xi}\\ 
& +\frac{\rho_+^2}{\mathcal{N}} \sum_{k=j}^{L}e^{-(L-k)/\xi}.
\end{split}
\label{tau_ij}
\end{equation}

Using eq.~(\ref{fluct_def}-\ref{tau_ij}) we get the following result for
the variance of the number of particles:
\begin{equation}
\begin{split}
\frac{\langle \Delta N \rangle}{L} & = \rho_- (1-\rho_-) \\
        &
+\frac{\delta}{L}\frac{(1-2\rho_-)e^{-1/\xi}}{1-e^{-(L+1)/\xi}}
        \left(\frac{1-e^{-L/\xi}}{1-e^{-1/\xi}}-Le^{-L/\xi}\right)\\
&+\frac{\delta^2}{L}\frac{e^{-2/\xi}(1-e^{-2L/\xi})-2
e^{-(L+3)/\xi}(1-e^{-L/\xi})}{(1-e^{-1/\xi})^2
(1-e^{-(L+1)/\xi})^2}\\
&-\delta^2 e^{-(L+1)/\xi}\frac{1+e^{-(L+1)/\xi}+L}{(1-e^{-(L+1)/\xi})^2}.
\end{split}
\end{equation}
The corresponding results in the high density phase can be obtained
by applying the particle hole symmetry of the TASEP.
On the coexistence line ($\alpha = \beta$) the localization length
diverges and the  fluctuations are given by: 
\begin{equation}
\frac{\langle \Delta N \rangle}{L}  =   \frac{\delta}{2}\,\left( 1 - 2\,\rho_- \right) + \left( 1 -
\rho_-\right) \,\rho_- + \delta^2\,\frac{\left( L - 4 \right)}{12}.
\end{equation}

\subsection{Comparison with simulation results}

\begin{figure}[h]
\centerline{\epsfig{figure=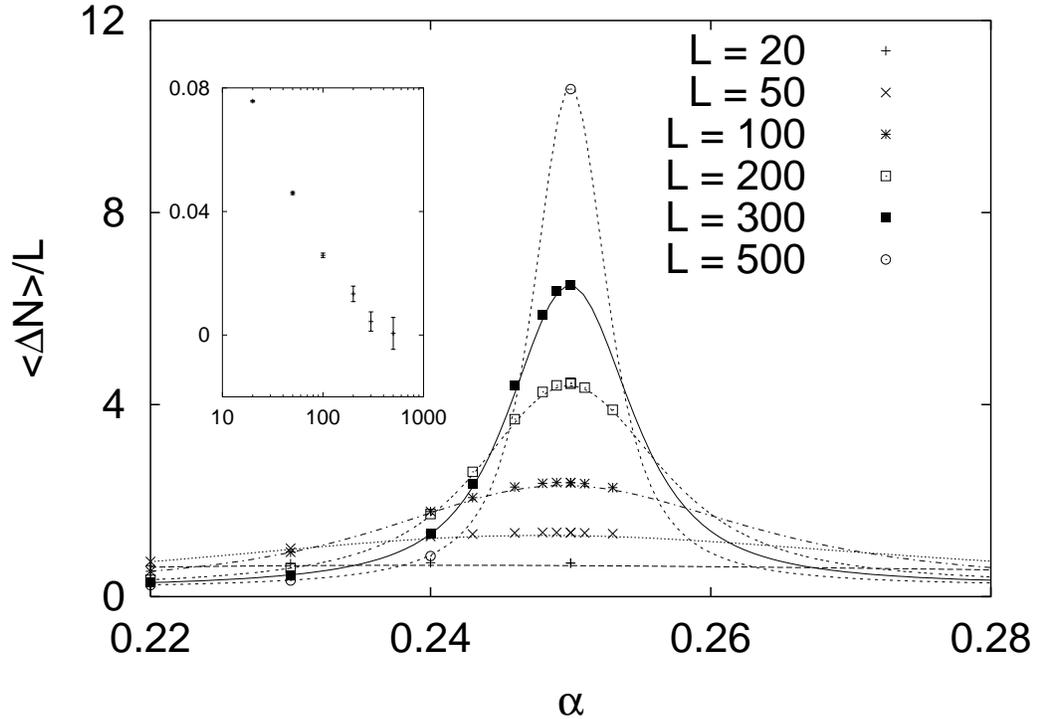,height=10cm}}
\caption{Domain-wall prediction {\sl vs.}  simulation results of the
particle number fluctuations $\langle \Delta N \rangle /L$. The
symbols correspond to simulations 
results of the finite chain. The predictions of the domain wall
picture are indicated by lines. We consider $\beta = 0.25$ for all
simulations. The inset shows the relative deviations $(\langle \Delta N 
\rangle - \langle \Delta N \rangle_{DW})/\langle \Delta N \rangle$ between 
simulations and domain-wall predictions. For $L\geq 300$ the devations are 
smaller than the numerical precision of our simulations. }
\label{fig_abtrans}
\end{figure}

Fig.~\ref{fig_abtrans} shows a comparison between the calculated values
and simulation results for the particle number fluctuations. The
results are for $\beta = 0.25$ and different values of $\alpha$ 
surrounding the first order transition at $\alpha=0.25$. Our
simulations start in the low density phase,
cross the coexistence line and end up in the high density
phase. The agreement between DW-theory and simulation
results is excellent for larger system sizes. For
small system sizes one observes small deviations from the
simulation results, which are presumably due to the finite width
of the shock.
Nevertheless, we stress the fact that the DW theory gives
correct results (i.e. within the accuracy of the simulations) already
for systems of the order of hundred sites.

\section{Non-stationary processes}

The results for the stationary particle number fluctuations presented
in the previous section - for which an exact solution exists -
have provided a test for the accuracy of the
domain-wall picture. But in contrast to the exact stationary solution
the domain wall picture also allows to characterize the dynamics of
the system in the transient regime. 

\subsection{Density profiles} 

In order to check the correctness of the phenomenological picture 
we investigated the time-dependent density profile interplaying
between two stationary states. We started from a stationary state in
the low density regime on the line $\alpha_0 + \beta_0 =1$. 
Indeed, on this line, the exact stationary solution has a simple 
product form, and the density profile is flat \cite{DEHP}.
This choice was made in order to keep the calculations simple,
and to generate easily a large number of independent
configurations.
Suddenly, we change the output rate $\beta$ such as to
{\sl(i)} lie on the coexistence line $\beta = \alpha = \alpha_0$
or {\sl (ii)} cross the transition line into the high density
phase so that $\beta < \alpha = \alpha_0$.
We take as the origin of time the moment when this change occurs.
A large number of independent simulations are performed in parallel.
At regular time intervals, an ensemble average of
the density profile over the independent simulations
is performed. After some time, a new stationary state is reached.

In our simulations, we chose as initial condition
$\alpha_0 = 0.25, \beta_0 = 0.75$.
At $t=0$, $\alpha = \alpha_0$ stays fixed, while $\beta$ is
varied to (i) $\beta = 0.25$, or (ii) $\beta = 0.2$, $0.1$, or $0.02$.
This change is instantaneous. The density profile is averaged every
100 time steps over 100,000 independent simulations, i.e. we took
100,000 arbitrary configurations from the initial stationary ensemble and
performed a simulation run for each initial configuration.

In the remaining part of this section we want to explain how this
setup can be described by means of the domain wall theory.
 
The chosen set of parameters ensures the absence of a shock in the
initial condition. By changing the right reservoir we introduced a
shock at position $L$ and time $t=0$. In the course of time  
the motion of the wall then follows a random walk described by
equations (\ref{eq_D1}-\ref{eq_D3}). 

In case (i) the random walk is symmetric, because $\alpha = \beta$ and
thus $D_+ = D_-$. An exact analytic solution 
for a random walk between two reflecting walls, starting
at r=0,
is known for all times \cite{schwarz_p75}:
\begin{eqnarray}
P(i,t) & = & \frac{1}{L+1} \left\{ 1 + \sum^{L+1}_{n=1}
\exp \left(-2Dt\left\{1-\cos\left[\frac{\pi n }{L+1}\right]\right\}
\right)\right. \nonumber\\
& & \left.\times \left[ \cos\left(\frac{i\pi n}{L+1}\right)
+ \cos\left(\frac{(i+1)\pi n}{L+1}\right)\right]\right\}.
\end{eqnarray}
At long times, the wall has a uniform probability to be located
anywhere in the system.

In case (ii), $D_+$ and $D_-$ are different. The exact analytical
solution of the asymmetric random walk with reflecting boundaries is also known
 \cite{Dud,schwarz_p75},
but for convenience we treated directly the discretized diffusion
equation 
\begin{eqnarray}
P(i,t+dt) & = & D_+ dt P(i-1,t) + D_- dt P(i+1,t)\nonumber\\
& &  + [1 - (D_+ + D_-) dt] P(i,t),
\end{eqnarray}
with equivalent expressions for the boundaries.
We took $dt$ small enough so that the results do not depend
on the chosen value.

Once the distribution $P(i,t)$ is known, the density profile
can be computed for any time
\begin{equation}
\rho(i,t) = \left( \sum_{n=0}^{i} P(n,t) \right) \rho_-
        + \left( \sum_{n=i+1}^{L} P(n,t) \right) \rho_+
\end{equation}
where $\rho_-$ and $\rho_+$ are the effective densities
of the left and right reservoirs, i.e. $\rho_- = \alpha$ and $\rho_+ =
1-\beta$ for the TASEP.

\begin{figure}[h]
\centerline{\psfig{figure=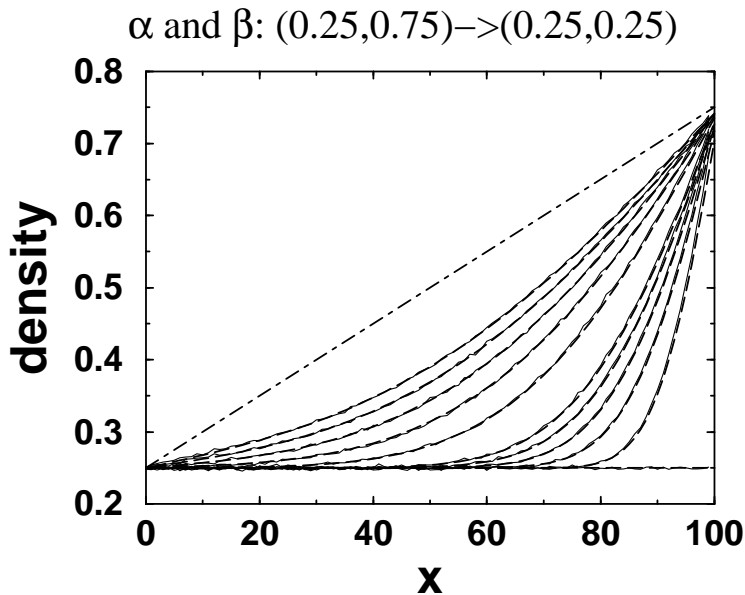,height=8cm}}
\caption{Domain-wall prediction (dashed lines) {\sl vs.}  simulation results
(solid lines) for the time dependent density profile when 
$\beta$ is suddenly changed at $t=0$ from $\beta=0.75$ to $0.25$.
The input rate is fixed $\alpha=0.25$ and the system size is $L=100$.
Each profile was averaged over 100000 independent simulations.
The dot-dashed line indicates the asymptotic linear density profile.
}
\label{fig_prof0.25}
\end{figure}

\begin{figure}[h]
\centerline{\psfig{figure=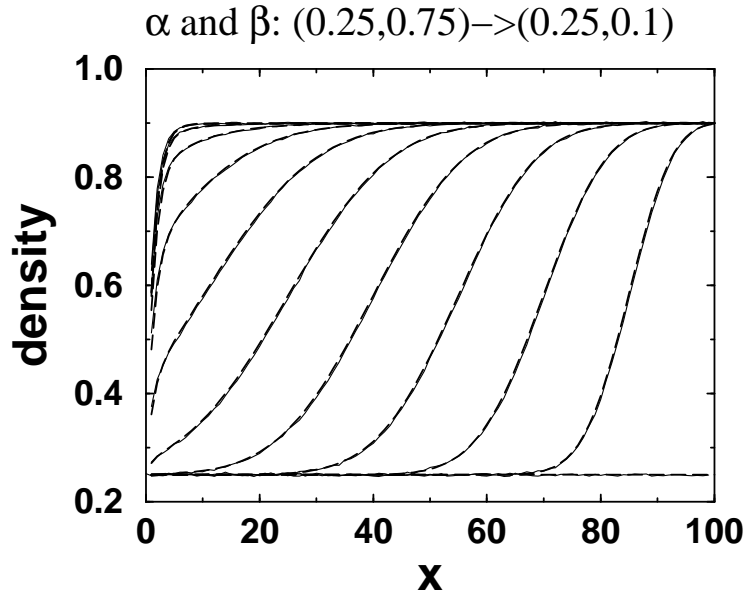,height=8cm}}
\caption{Domain-wall prediction (dashed lines) {\sl vs.}  simulation results
(solid lines) for the time dependent density profile when 
$\beta$ is suddenly changed at $t=0$ from $\beta=0.75$ to $0.1$.
The input rate is fixed $\alpha=0.25$.
Profiles are plotted every 100 time steps.
Each profile was averaged over 100000 independent simulations.
The system size is $L=100$.
}
\label{fig_prof0.1}
\end{figure}

The comparison with the simulation results (see figs.
\ref{fig_prof0.25} and \ref{fig_prof0.1}) shows an excellent
agreement. In case (i), the domain wall undergoes a diffusive motion
and the density profile modification scales as $\sqrt{t}$.
To make the picture more clear, beyond $t=500$,
we show only the profiles every 500 time steps.
In the long time limit, the domain wall is delocalized
over the whole system and the density profile is linear.

\begin{figure}[h]
\centerline{\psfig{figure=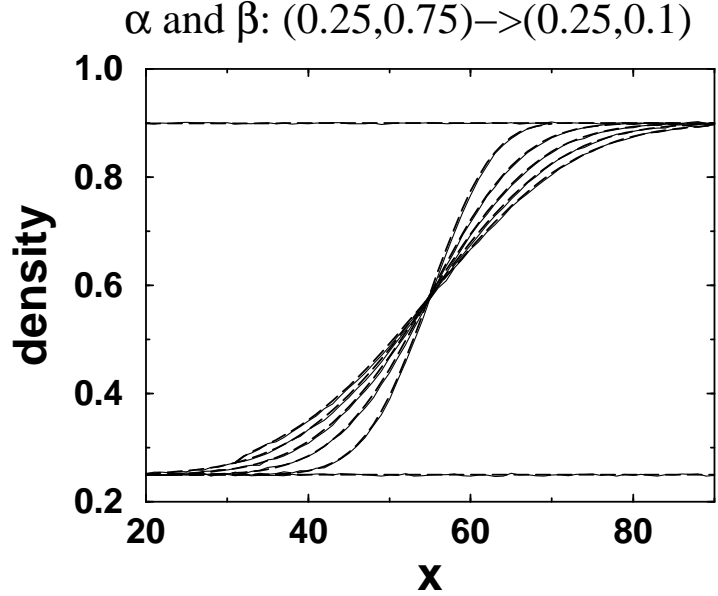,height=8cm}}
\caption{
Same figure as \protect{\ref{fig_prof0.1}}, but with
the density profiles translated by a multiple of
100 V.
}
\label{fig_prof0.1bis}
\end{figure}

In case (ii), the wall has a non vanishing drift velocity,
and the shift of the density profile is
linear in time, at least as long as boundaries are far enough.
This is illustrated in fig. \ref{fig_prof0.1bis} where
the first density profiles
(for $100 \leq t \leq 500$)
are translated by a multiple of 100 V. The 
averaged wall velocity $V$ is given by eq. (\ref{V-wall}). 
Of course, as the wall position distribution evolves from
a Dirac function into a larger and larger gaussian, the
density profile flattens as time increases.
At long times, the wall is located near the left boundary,
and the density profile is flat, apart from an exponential
boundary layer at the entrance of the system
(fig. \ref{fig_prof0.1}).

\begin{figure}[h]
\centerline{\psfig{figure=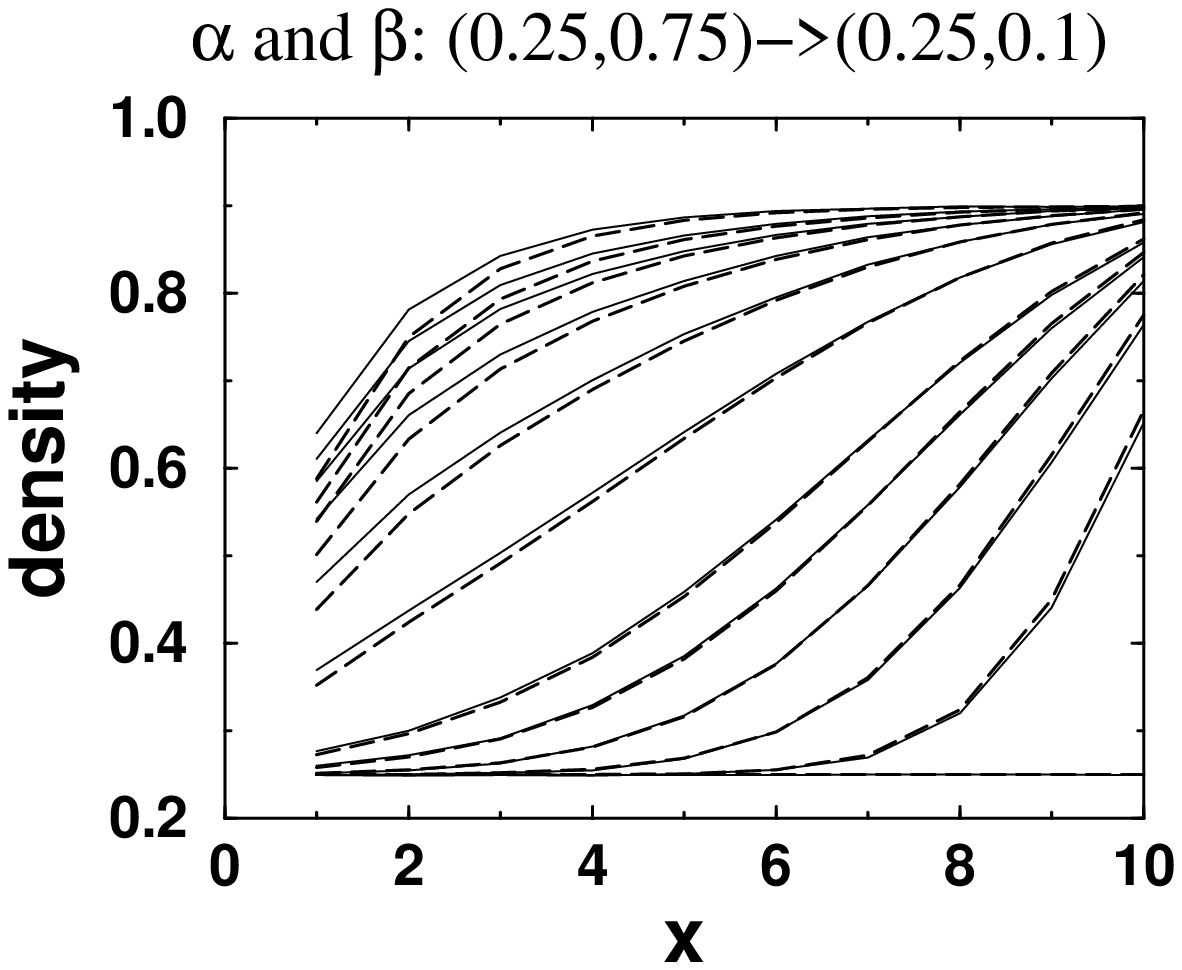,height=8cm}}
\caption{Domain-wall prediction (dashed lines) {\sl vs.}  simulation results
(solid lines) for the time dependent density profile when 
$\beta$ is suddenly changed at $t=0$ from $\beta=0.75$ to $0.1$.
The input rate is fixed $\alpha=0.25$.
Profiles are plotted at time $0$, $5$, $10$, $15$, $20$, $25$, $28$, $31$,
$34$, $37$, $40$, and $500$.
Each profile was averaged over 100000 independent simulations.
The system size is $L=10$.
}
\label{fig_l10}
\end{figure}

For small systems, the agreement between domain wall
theory and simulations is not so good, as expected. 
However, even for a system of size $L=10$, the 
domain wall results still indicate quite well 
the non-stationary behavior, as long as the wall
is far enough from the left boundary (fig \ref{fig_l10}).
Surprinsingly, the agreement is much better at the
beginning, when the wall is still near the right boundary.
It could be linked to the fact that in the initial
condition, as the density profile is flat, the domain
wall approximation (with a wall on the right of the
system) is exact.

\subsection{Fluctuations in the particle number}

\begin{figure}[h]
\centerline{\psfig{figure=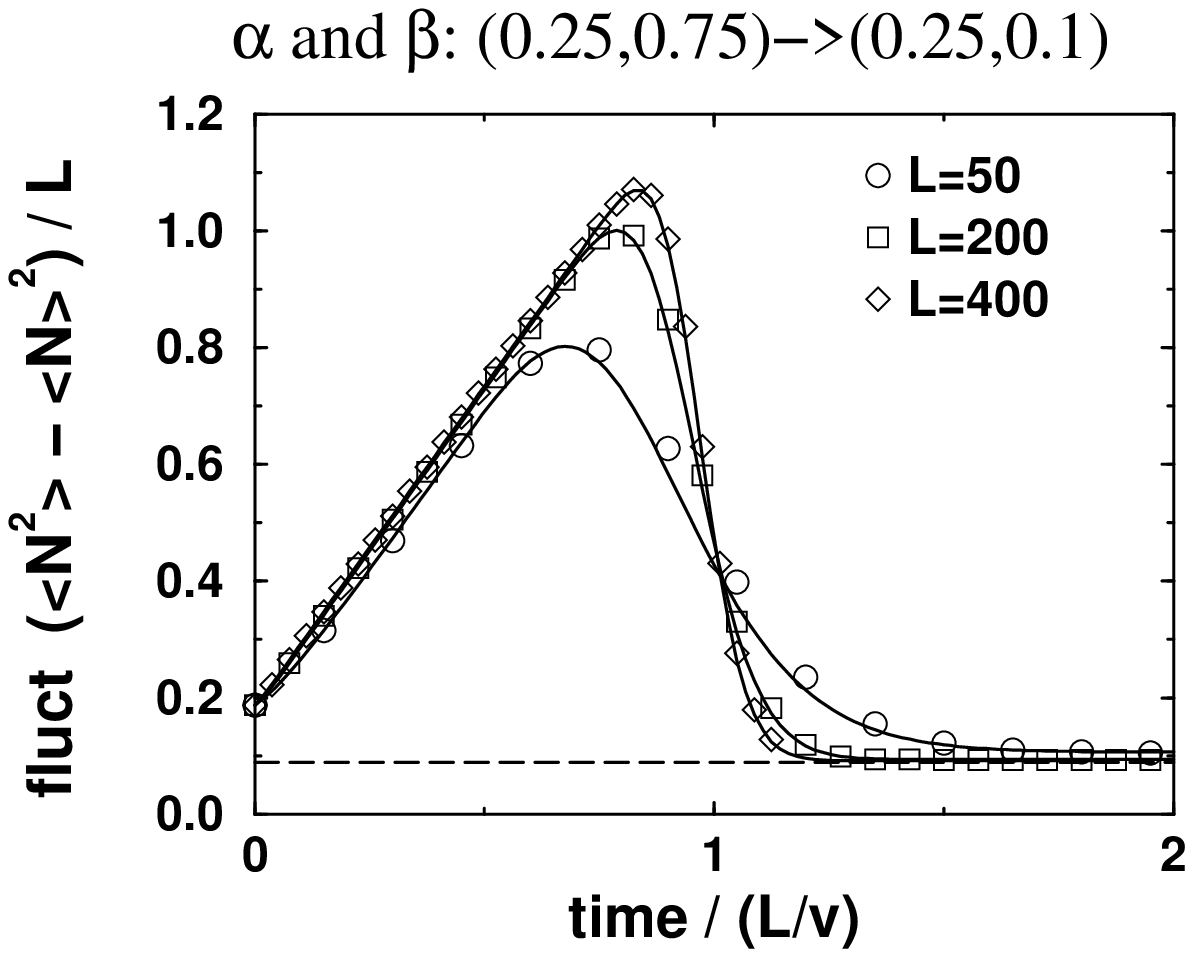,height=8cm}}
\caption{Fluctuations in the total number of particles
as a function of time (rescaled by $L/V$). The simulations (symbols) are
compared with the domain wall estimate
(black continuous lines) with excellent agreement.
The dashed line indicates the value $\rho_+(1-\rho_+) = 0.09$.
}
\label{fig_flu}
\end{figure}

So far we focused on density profiles.
Now we shall show that the domain wall picture allows
also to predict the fluctuations in the total number of particles
in a dynamical regime.
 From the calculation in the previous section, the time dependent 
probability distribution $P(i,t)$ is known.
The fluctuations are evaluated from formulas similar to
(\ref{Nmoy}-\ref{tau_ij}), but now we replace the stationary distribution
$e^{-(L-i)/\xi}$ by $P(i,t)$.
This yields
\begin{equation}
\langle N \rangle(t) = L \rho_- + (\rho_+ - \rho_-) \sum_{k=0}^{L} (L-k) P(k,t)
\end{equation}
and
\begin{equation}
\begin{split}
\langle \tau_i\tau_j \rangle(t) & = \frac{\rho_-^2}{\mathcal{N}} \sum_{k=0}^{i-1}
P(k,t) +
\frac{\rho_-\rho_+}{\mathcal{N}}\sum_{k=i}^{j-1}P(k,t)\\ 
& +\frac{\rho_+^2}{\mathcal{N}} \sum_{k=j}^{L}P(k,t).
\end{split}
\label{tau_ij2}
\end{equation}
Using these expressions with the definition (\ref{fluct_def}),
the variance of the number of particles can be computed numerically
for any time, and compared with the direct simulation results.
Fig. \ref{fig_flu} presents such a comparison in case (ii),
i.e. for a final value $\beta=0.1$
Time has been divided by the average time needed for
the wall to cross the system, i.e. $L / V$ where $V$
is given by equation (\ref{V-wall}).
The agreement is excellent for all sizes, though for $L=50$,
some small finite size effects are visible.

In the early linear stage, the wall has not yet reached the
left boundary, and the curves superimpose for all sizes.
When the wall arrives near the left boundary, the width of the
probability distribution $P(i,t)$ scales as $\sqrt{L}$.
Thus the typical time during which domain walls reach the
left boundary in the different realizations of the system
also scales as $\sqrt{L}$.
As time has been rescaled by $L$ in fig. \ref{fig_flu},
this transient separating the linear growth from the
final stationary state tends to a sharp jump as $L$
becomes large.

In the final stationary state, the fluctuations within the high density
phase dominate. They can be estimated from the mean field
expression $\rho_+(1-\rho_+) = (1-\beta)\beta = 0.09$.

\section{Conclusion} 

Our results have shown that many features of the TASEP with open
boundary conditions 
can be understood by analyzing the domain wall dynamics.
For the stationary state, where the DW theory is known to give exact
results in the limit of large $L \to \infty$ \cite{Belitzky}, we
calculated the 
fluctuations of the particle number for chains of various
sizes. As being the integrated two-point correlation functions the 
fluctuations of the particle number are a sensitive test for 
the accuracy of the DW theory. We found that the predictions
of the DW theory are numerically indistinguishable from the simulation
results for  $L\gtrsim 100$.

But beyond that we also found an excellent agreement in the transient
regime. This is of particular interest, because the transient regime
is out of scope of exact solution. The perfect agreement (up to the
precision of our measurements) between
simulation results and DW theory shows that the domain wall motion
determines exclusively the relaxation processes, i.e. the coupling to 
the particle reservoirs is instantaneous. For future work it will be
interesting to see whether other relaxation mechanisms are of
importance if the domains are not of a simple product measure. 

In several other studies it has been shown that the DW approach
correctly predicts the phase diagram for a large number of
models and update schemes \cite{Popkov00,Pigorsch,Kaidiss}. In
general cases, where an exact solution of the process is not possible,
one has to establish the coupling to the chain and the flow density relations
numerically. However, also in this case the domain wall interpretation
is useful: First of all the parameter space is reduced drastically
(e.g. the bulk density corresponding to a particular input prescription
can be obtained from a {\em single} simulation run with free right
boundary),
and second it allows for a better characterization of the physical 
behavior \cite{Popkov00}. Let us illustrate the latter point by the
example of the 
Nagel-Schreckenberg model for traffic flow with maximal velocities
$v_{max}>1$ \cite{NaSch}. In 
this case one has several possibilities to implement the in- and output of the
particles. This choice can produce even qualitatively  
different results for the phase diagram.
E.g. it is necessary to apply a reservoir that can achieve the full
capacity of the chain for a given model, in order to observe the
maximum current phase (see \cite{Chey} for a counter-example). While
for the TASEP the natural input of particles on the first site of the
chain fulfills this prerequisite,  
this is more subtle for particles which can move with higher
velocities \cite{Popkov00}. The difficulties in finding a suitable
particle reservoir 
can be circumvented using the basic idea of the domain wall theory for
the description of the particle reservoir. Instead of describing the
capacity of the reservoirs in terms of in- and output probabilities 
we recommend  to describe a 
particle reservoir by its effective density, i.e. by the bulk density
which corresponds to a specific parameter combination, and to calculate
the phase diagram only thereafter.

Summarizing, we have shown that the DW theory is able to reproduce the
behavior of the TASEP with open boundary conditions to a large
extend, and we have used it to obtain new results for non-stationary
flows. We expect that our results can be generalized to a wide class
of models for particle transport.

{\bf Acknowledgments}: We acknowledge fruitful discussions with
G. Sch\"utz, B. Derrida, and A. Schadschneider.  
L.~S. acknowledges support from the Deutsche
Forschungsgemeinschaft under Grant No. SA864/1-2. 

\bibliographystyle{unsrt}

\begin{thebibliography}{99}


\bibitem{spohn} H. Spohn, {\em Large Scale Dynamics of Interacting
Particles}, (Springer, Berlin, 1991) 
 
\bibitem{sz} B. Schmittmann and R.K.P. Zia, in: {\em Phase Transitions
    and Critical Phenomena}, Vol.~17, eds. C. Domb and J.L. Lebowitz 
  (Academic Press, New York, 1995)

\bibitem{gunter} G.M.~Sch\"utz, in: 
{\em Phase Transitions and Critical Phenomena}, Vol. 19, eds. C. Domb
and J.L. Lebowitz (Academic Press, New York,  2000)

\bibitem{review}
D.~Chowdhury, L.~Santen and A.~Schadschneider,
Physics Reports {\bf 329}, 199 (2000)

\bibitem{MacD69}
J.T.~MacDonald and J.H.~Gibbs, Biopolymers {\bf 7}, 707 (1969)

\bibitem{Krug91}
J.~Krug, Phys. Rev. Lett.~{\bf 67}, 1882 (1991)


\bibitem{Liggett} T.M.~Liggett, Trans. Amer. Math. Soc. {\bf 179}, 433 (1975)

\bibitem{Derr92} B.~Derrida , E.~Domany and D. Mukamel, J. Stat. Phys.,
667 (1992) 

\bibitem{Gunter93}
 G.M.~Sch\"utz and E.~Domany, J.~Stat.~Phys.~{\bf 72}, 277 (1993)


\bibitem{DEHP} B.\ Derrida, M.R.\ Evans, V.\ Hakim, and V.\ Pasquier, 
J. Phys. A~{\bf 26}, 1493 (1993)


\bibitem{Kolo98}
A.B.~Kolomeisky, G.M.~Sch\"utz, E.B.~ Kolomeisky, and  J.P.~Straley,
J.Phys.~A {\bf 31}, 6911 (1998)

\bibitem{Belitzky} V. Belitzky and G.M. Sch\"utz, to be published

\bibitem{Popkov99} V. Popkov and G.M. Sch\"utz, 
Europhys.\ Lett. {\bf 48}, 257 (1999)

\bibitem{Popkov00} V.~Popkov, L.~Santen, A.~Schadschneider and
G.M.~Sch\"utz, J. Phys. A {\bf 34}, L45 (2001) 

\bibitem{Dud} M. Dudzinski and G.M.~Sch\"utz, J.~Phys. A {\bf 33}, 8351 (2000)

\bibitem{Derrida} B. Derrida, J.L. Lebowitz and E.R. Speer,
preprint, cond-mat/0105110.

\bibitem{schwarz_p75} M. Schwartz and D. Poland, J. Chem. Phys., {\bf
63}, 1 (1975) 

\bibitem{Pigorsch} 
  C Pigorsch and Sch\"utz, J.~Phys.~ A {\bf 33}, 7919 (2000)

\bibitem{Kaidiss} 
 K. Klauck, PhD thesis, Cologne (2000)

\bibitem{Chey} S. Cheybani, J. Kertesz and M. Schreckenberg,
Phys. Rev. E {\bf 63}, 016108 (2001)    

\bibitem{NaSch} K.~Nagel and M.~Schreckenberg,
J. Physique I, {\bf 2}, 2221 (1992)

\end{thebibliography}

\end{document}